\documentclass[conference]{IEEEtran}

\usepackage[utf8]{inputenc}
\usepackage[T1]{fontenc}
\usepackage[english]{babel}

\usepackage{graphicx}
\usepackage{float}
\usepackage{subfigure}
\usepackage{mathtools}
\usepackage{booktabs}
\usepackage{hyperref}
\usepackage{cleveref}
\usepackage[nolist,nohyperlinks]{acronym}

\graphicspath{{images/}}

\title{
EEG Interpretation Across Chant Listening:
A Single-Subject Pilot Investigation Using Spectral and Functional Connectivity Analysis
}

\author{

\IEEEauthorblockN{
Prerna Singh\textsuperscript{1},
Aishwarya Ghosh\textsuperscript{2},
Neelam Sinha\textsuperscript{1},
Deepti Navaratna\textsuperscript{2*}\thanks{Corresponding author: [deepti.navaratna@nias.res.in](mailto:deepti.navaratna@nias.res.in)}
}

\IEEEauthorblockA{\textsuperscript{1}
Centre for Brain Research (CBR), Indian Institute of Science (IISc), Bengaluru, India\
}

\IEEEauthorblockA{\textsuperscript{2}
National Institute of Advanced Studies (NIAS), Bengaluru, India\\ *email: deepti.navratna@nias.res.in
}

}

\begin{document}

\maketitle

\begin{abstract}

This technical report presents an EEG-based investigation of neural activity across five auditory conditions: Resting State (RS), Shiv Tandav Stotra (STS), Mahasudarshan Mantra (MM), Aum Chant, and Tanpura Listening. EEG recordings acquired from a healthy 5-year-old participant were analyzed using spectral power estimation and functional connectivity measures based on the weighted Phase Lag Index (wPLI). Spectral analysis revealed condition-specific modulation of neural oscillatory activity, with STS listening producing the highest relative power across multiple frequency bands, particularly within the beta range. Functional connectivity analysis demonstrated distinct network organizations across conditions. STS listening exhibited the strongest and most widespread connectivity pattern, characterized by prominent long-range interactions among frontal, temporal, parietal, and occipital regions. Tanpura listening generated a dense yet balanced connectivity network, while Aum listening showed moderate distributed connectivity. In contrast, MM and resting-state conditions displayed comparatively weaker and more localized network organization. These preliminary findings suggest that different chant-listening conditions engage distinct neural mechanisms involving both cortical activation and large-scale neural synchronization. The study establishes a methodological framework for future investigations examining the role of culturally relevant auditory interventions in cognitive development, neuroeducation, and child-centered neuroscience research.

\end{abstract}


\section{Introduction}

India is undergoing a major transformation in its educational ecosystem through the implementation of the National Education Policy (NEP) 2020, which emphasizes holistic, multidisciplinary, and developmentally aligned learning. Despite this progress, a critical gap remains in our understanding of how Indian children learn, develop, and respond to educational interventions from a neurobiological perspective.

Globally, advances in educational neuroscience have demonstrated that cognitive abilities such as attention, memory, executive function, language processing, and self-regulation are closely associated with the maturation of neural networks during childhood \cite{Posner2007,Johnson2011}. However, most existing evidence originates from Western populations and may not adequately represent the unique sociocultural, linguistic, environmental, and educational contexts of India.

Several reports have highlighted the need for early cognitive assessment and intervention among Indian children. The Annual Status of Education Report (ASER) demonstrated that significant learning gaps emerge even before formal schooling begins, particularly in rural populations \cite{ASER2019}. Cognitive skills such as pattern recognition, spatial reasoning, sequencing, and working memory have been shown to predict later literacy and numeracy outcomes \cite{Rao2014,Das1989}. Furthermore, Mishra emphasized the importance of developing culturally relevant cognitive and developmental frameworks rather than relying exclusively on Western developmental models \cite{Mishra1998}.

Recent developments in neuroeducation have highlighted the benefits of neuroscience-informed interventions in improving learning outcomes. Brain-based learning approaches have demonstrated improvements in conceptual understanding, academic performance, and cognitive engagement across different educational settings \cite{Varghese2016,Sharma2024}. Furthermore, interventions targeting executive function, multisensory processing, and cognitive enrichment have shown promise in children with developmental and learning difficulties \cite{Devaraj2018,Hada2022,Singh2021}.

One promising but relatively underexplored avenue involves meditation, chanting, and rhythmic auditory stimulation as potential tools for cognitive enhancement and neural regulation. Previous EEG and neuroimaging studies have shown that auditory exposure to rhythmic vocalizations, mantra sounds, and chant stimuli can modulate neural oscillations and large-scale brain networks involved in attention, sensory integration, and emotional processing \cite{Nozaradan2012,Lomas2015, cahn2006meditation}. Studies examining AUM chant listening and related auditory mantra paradigms have reported changes in cortical synchronization, alpha--theta activity, and autonomic regulation \cite{Kalyani2011}.Similar findings have been reported during mantra meditation paradigms, where increased alpha coherence and altered fronto-parietal communication were observed during sustained chanting and focused auditory attention \cite{hinterberger2014meditation}. Rhythmic auditory stimulation may further facilitate neural entrainment mechanisms that support attention and cognitive engagement \cite{Lomas2015}.

The long-term vision of the present research program is to investigate whether neurophysiological signatures derived from EEG can contribute toward the development of a brain-based developmental schema capable of characterizing diverse cognitive trajectories in children. Such a framework may eventually help identify children exhibiting advanced cognitive abilities, typical developmental patterns, or developmental vulnerabilities requiring early intervention. Early detection is particularly relevant for neurodevelopmental conditions such as autism spectrum disorder (ASD), attention-deficit/hyperactivity disorder (ADHD), and specific learning disorders, where timely intervention has been shown to significantly improve outcomes \cite{Jeste2015,Uddin2017}.

Electroencephalography (EEG) offers a non-invasive, cost-effective, and temporally precise method for monitoring neural oscillations and functional brain connectivity. EEG biomarkers have been widely used to investigate cognitive development, attention networks, learning processes, and neurodevelopmental disorders \cite{Klimesch1999,Posner2007}. As a preliminary step toward this broader objective, the present pilot study examines neural responses to a range of auditory conditions, including Resting State (RS), Shiv Tandav Stotra (STS), Mahasudarshan Mantra (MM), Aum Chant Listening, and Tanpura Listening. Understanding how these practices modulate neural dynamics may provide preliminary evidence regarding their potential utility as cognitive enrichment or therapeutic tools for supporting healthy neurodevelopment and neurorehabilitation in children.

\section{Experimental Methodology}

\subsection{EEG Acquisition}

To investigate the neural effects of chant listening and auditory stimulation, electroencephalography (EEG) recordings were acquired from a healthy 5-year-old participant using the NeuroMax digital EEG system (Medicaid Systems, India). The recordings were obtained using 19 scalp electrodes positioned according to the International 10--20 electrode placement standard, enabling the acquisition of neural activity from frontal, central, temporal, parietal, and occipital brain regions.

The experimental paradigm consisted of five distinct auditory conditions designed to investigate neural responses to chant listening and sound-based stimulation:

\begin{enumerate}
    \item Rest (Baseline)
    \item STS Chant Listening
    \item Mahasudarshan Mantra Listening
    \item AUM Chant Listening
    \item Tanpura Listening
\end{enumerate}

\begin{figure}[H]
\centering
\includegraphics[width=0.95\linewidth]{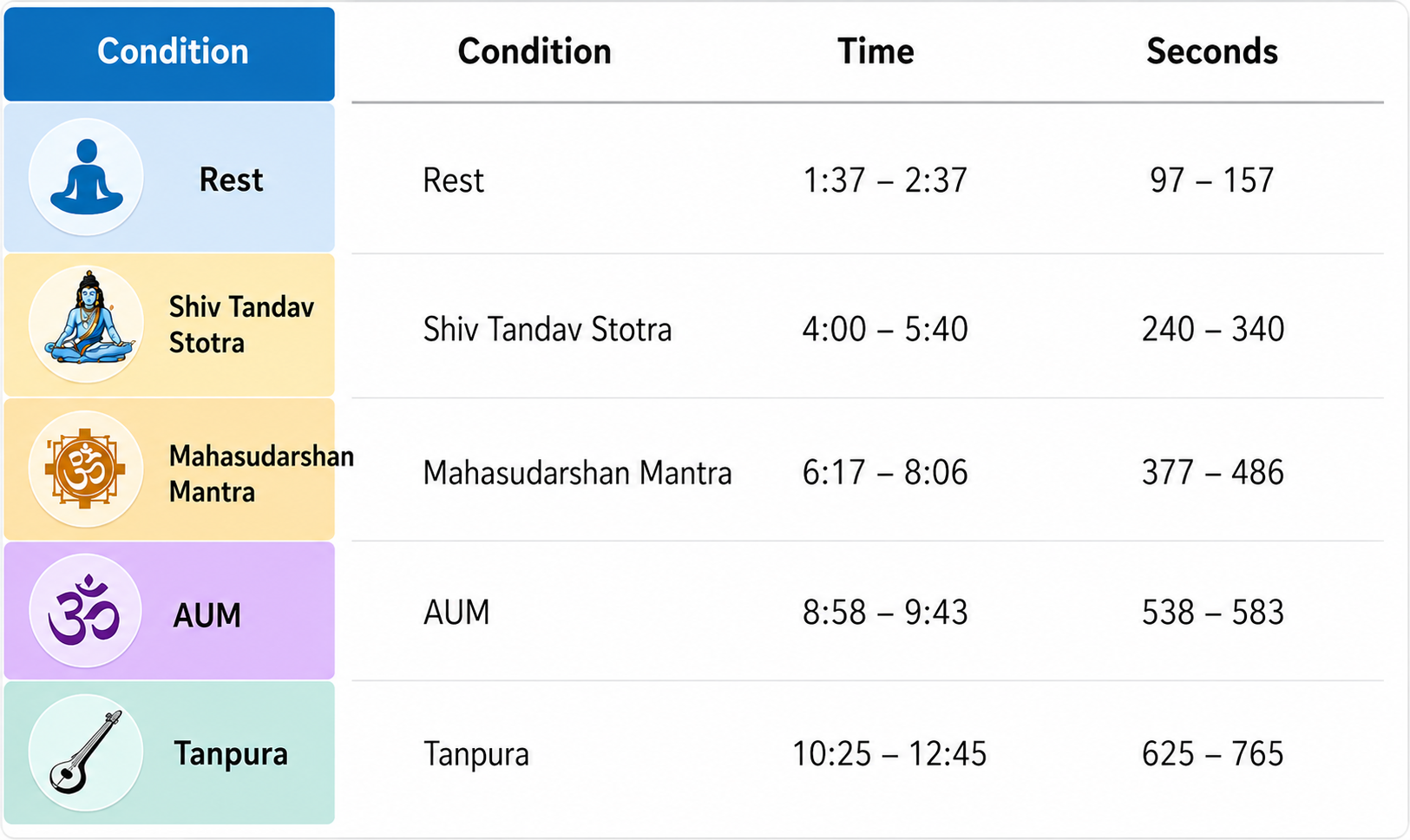}
\caption{Timeline of experimental conditions used during EEG acquisition.}
\label{fig:timeline}
\end{figure}

For each condition, continuous EEG data were recorded for a duration of approximately 2 minutes while the participant listened attentively to the corresponding auditory stimulus. Given the young age of the participant (5 years), short transition periods were incorporated between consecutive conditions to ensure comfort, maintain engagement, and minimize movement-related artifacts. The order and duration of the recording segments were selected to obtain stable neural activity representative of each listening condition while accommodating the participant's attention span. Figure~\ref{fig:timeline} illustrates the sequence of experimental conditions employed during EEG acquisition.

The recorded EEG signals were subsequently subjected to a standardized preprocessing pipeline, followed by spectral and functional connectivity analyses to characterize condition-specific neural oscillatory dynamics and large-scale network organization.

\subsection{EEG Preprocessing}

Raw EEG recordings contain physiological and environmental artifacts that can significantly affect spectral and connectivity analyses. Therefore, a standardized preprocessing pipeline was implemented using EEGLAB.

Initially, EEG recordings were imported into EEGLAB and channel labels were verified according to the International 10--20 system. Condition-specific segments corresponding to the five experimental states were extracted from the continuous recordings.

To improve stationarity and facilitate subsequent analyses, the extracted segments were divided into non-overlapping epochs of one-second duration. EEG signals were then filtered to isolate the alpha frequency band (8--12 Hz), which is known to play a critical role in attention, relaxation, and meditative processing.

Common Average Referencing (CAR) was subsequently applied to reduce reference-related bias and improve spatial representation of cortical activity. Independent Component Analysis (ICA) was then employed to identify and remove artifacts arising from eye blinks, eye movements, muscle activity, and external noise sources.

\begin{figure*}
 
\centering
\includegraphics[width=0.95\linewidth]{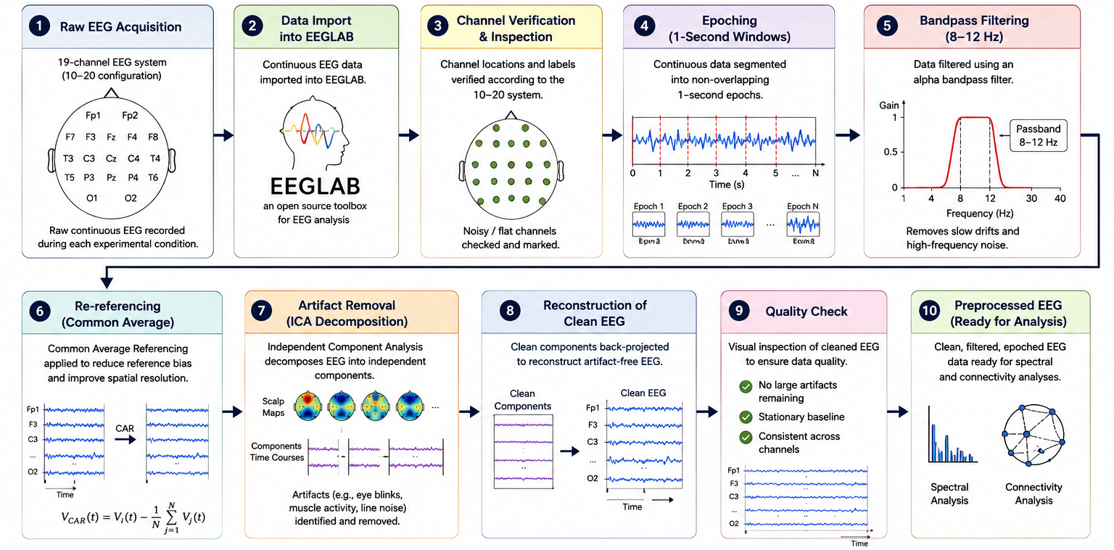}
\caption{EEG preprocessing and analysis pipeline employed in the study.}
\label{fig:pipeline}

\end{figure*}

Figure \ref{fig:pipeline} summarizes the complete EEG processing workflow.

\subsection{Spectral Analysis}

Spectral analysis was performed to investigate condition-dependent changes in neural oscillatory activity. Power Spectral Density (PSD) was estimated for each EEG epoch and subsequently averaged across epochs within each condition.

Relative band power was computed for four canonical EEG frequency bands:

\begin{itemize}
\item Delta (0.5--4 Hz)
\item Theta (4--8 Hz)
\item Alpha (8--12 Hz)
\item Beta (13--30 Hz)
\end{itemize}

The resulting power estimates enabled quantitative comparison of oscillatory dynamics across different meditative states. In addition, channel-wise alpha power distributions were examined to identify regional changes associated with attention, relaxation, and neural synchronization.

\subsection{Functional Connectivity Analysis}

Functional connectivity analysis was conducted to evaluate large-scale communication patterns between cortical regions.

Two complementary phase-based connectivity metrics were employed:

\begin{itemize}
\item Phase Lag Index (PLI)
\item Weighted Phase Lag Index (wPLI)
\end{itemize}

PLI quantifies the consistency of non-zero phase relationships between EEG signals, while wPLI reduces the influence of volume conduction and noise, providing a more robust estimate of true functional interactions.

Connectivity matrices were computed for each experimental condition. To emphasize the most significant neural interactions, only the top 10

\section{Results and Interpretation}

\subsection{Relative Band Power Analysis}

\begin{figure*}
\centering
\includegraphics[width=0.9\linewidth]{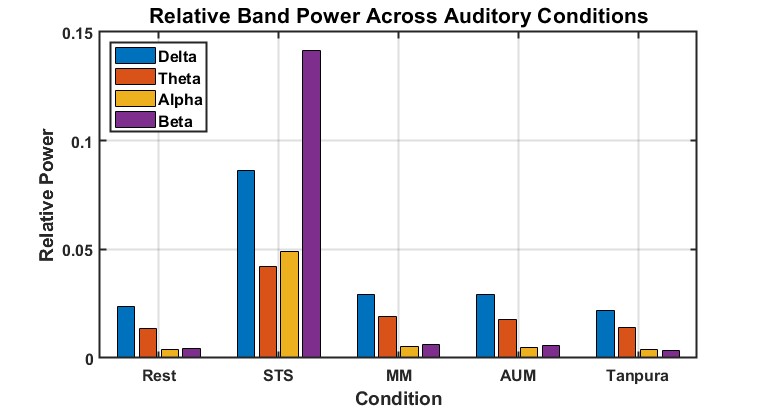}
\caption{Comparison of normalized spectral band power in the delta, theta, alpha, and beta frequency bands during Resting State (RS), Shiv Tandav Stotra (STS), Mahasudarshan Mantra (MM), AUM Chant, and Tanpura listening. STS exhibited the highest power across multiple frequency bands, suggesting enhanced cortical activation and attentional engagement, whereas AUM and Tanpura conditions demonstrated comparatively balanced oscillatory activity.}
\label{fig:bandpower}
\end{figure*}

Figure \ref{fig:bandpower} presents the relative spectral power observed across all experimental conditions.

The STS listening condition demonstrated the highest power across all frequency bands, particularly within the beta range. Elevated beta activity is often associated with increased attentional engagement, cognitive processing, and active mental involvement. In addition, STS listening showed increased delta, theta, and alpha activity, suggesting widespread cortical activation.

The resting condition exhibited the lowest overall spectral power and served as the baseline neural state. Aum chant and Tanpura conditions demonstrated moderate delta and theta activity while maintaining relatively low beta power. These observations indicate reduced cortical arousal and a more relaxed neural state.

MM exhibited intermediate spectral characteristics, showing slightly increased power relative to rest but lower activation compared to STS listening.

Overall, the spectral findings suggest that STS listening induces strong neural activation, whereas Aum chant and Tanpura listening conditions are associated with calmer and potentially more introspective cognitive states.

\subsection{Alpha Power Distribution}

\begin{figure*}
\centering
\includegraphics[width=0.85\linewidth]{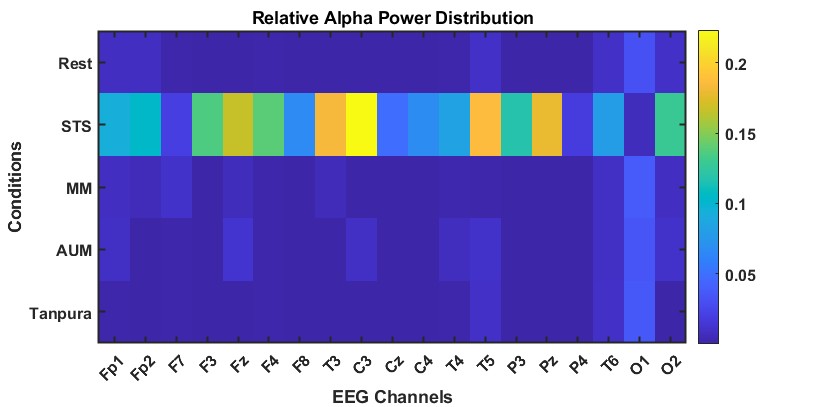}
\caption{Spatial distribution of alpha-band power across experimental conditions.
Channel-wise alpha power heatmaps illustrating regional variations in cortical oscillatory activity during different auditory conditions. Enhanced frontal and central alpha activity was observed during STS listening, while AUM, MM, and Tanpura conditions showed more distributed and moderate alpha-band modulation.}
\label{fig:alpha}
\end{figure*}

Alpha oscillations are frequently associated with attentional control, relaxation, and meditative processing. Figure \ref{fig:alpha} illustrates channel-wise alpha power distributions for all conditions.

A pronounced increase in alpha power was observed during STS listening, particularly over frontal electrodes (F3, Fz, and F4). Additional alpha enhancement was observed in central and parietal regions, indicating widespread recruitment of cortical networks.

The resting condition demonstrated weak posterior alpha activity, consistent with classical resting-state EEG observations. In contrast, Aum, MM, and Tanpura exhibited relatively uniform alpha distributions with lower overall amplitudes.

Interestingly, despite showing the strongest connectivity patterns, listening to Aum did not demonstrate the highest alpha power. This suggests that the mental state induced by Aum listening may be characterized more by enhanced neural synchronization than by simple increases in oscillatory amplitude.

\subsection{Spectral Characteristics}

\begin{figure*}
    \centering
\includegraphics[width =0.8\linewidth]{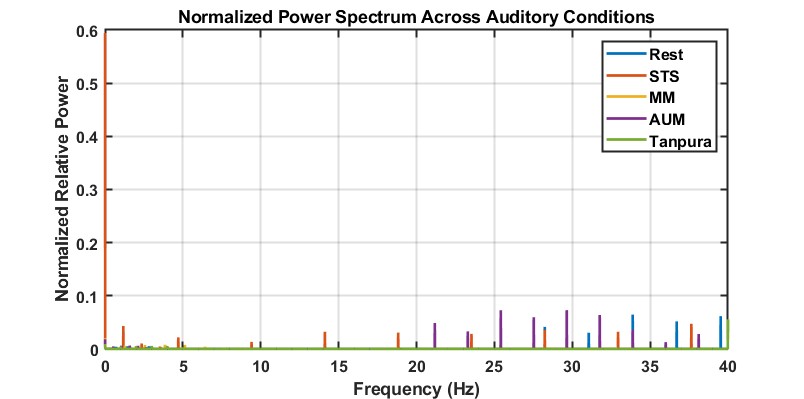}
\caption{Normalized EEG power spectra during auditory stimulation.
Average normalized power spectral density profiles for Resting State (RS), Shiv Tandav Stotra (STS), Mahasudarshan Mantra (MM), AUM Chant, and Tanpura listening. Distinct spectral signatures were observed across conditions, with STS exhibiting pronounced broadband activation and stronger alpha–beta spectral peaks compared with the other auditory stimuli.}
\label{fig:spectrum}

\end{figure*}

The normalized power spectra are shown in Figure \ref{fig:spectrum}.

Distinct peaks were observed within the alpha (8--12 Hz) and beta (20--35 Hz) frequency ranges. STS listening exhibited strong broadband activation with pronounced spectral peaks, indicating widespread engagement of neural populations.

Aum listeing demonstrated selective enhancement in higher frequencies, whereas Tanpura exhibited a smoother spectral profile characteristic of relaxed neural processing. Resting-state activity showed comparatively weaker and more distributed spectral content.

\subsection{Functional Connectivity Analysis}

\begin{figure*}
\centering
\includegraphics[width=0.85\linewidth]{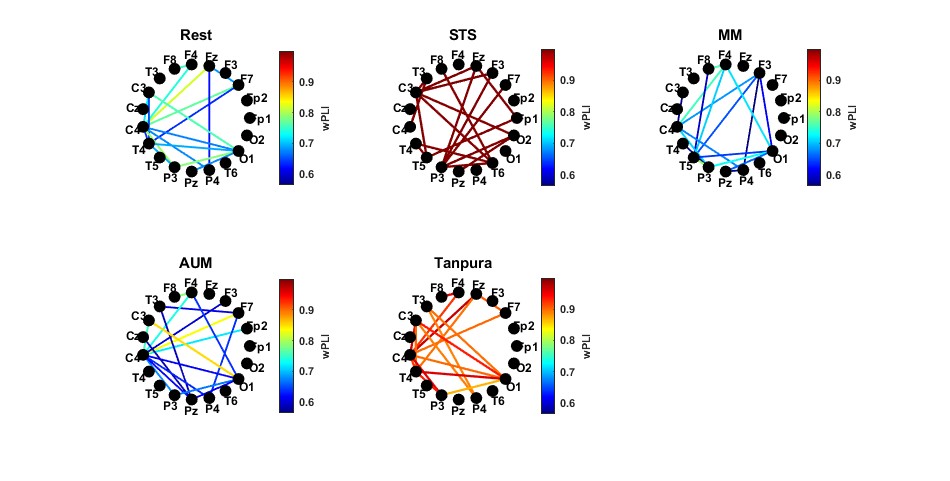}
\caption{Functional connectivity networks obtained from the top 10\% strongest weighted Phase Lag Index (wPLI) connections during Rest, STS (Shiv Tandav Stotra (STS), Mahasudarshan Mantra (MM), AUM, and Tanpura listening conditions. Edge colors represent connectivity strength (wPLI), with warmer colors indicating stronger functional coupling between brain regions. STS demonstrated the most extensive and strongest long-range connectivity, while AUM and Tanpura exhibited organized distributed networks, indicating condition-specific modulation of functional brain connectivity.}
\label{fig:connectivity}
\end{figure*}

Functional connectivity networks derived from wPLI are shown in Figure \ref{fig:connectivity}.

\subsection{Functional Connectivity Analysis}

Functional connectivity networks derived from the top 10\% strongest weighted Phase Lag Index (wPLI) connections are illustrated in Figure \ref{fig:connectivity}. The color scale represents connection strength, with warmer colors indicating stronger phase synchronization between electrode pairs.

The resting condition exhibited moderate connectivity characterized by a combination of local and long-range interactions. Although several distributed connections were observed, the overall network organization appeared less structured than the auditory listening conditions.

STS listening demonstrated the strongest and most extensive connectivity pattern among all conditions. Dense long-range interactions connecting frontal, temporal, parietal, and occipital regions were evident, with several connections exhibiting high wPLI values. These findings suggest enhanced large-scale neural synchronization and widespread functional integration during rhythmic chant listening.

MM listening produced a comparatively sparse and selective connectivity network. The strongest interactions were confined to a smaller subset of cortical regions, indicating a more localized pattern of neural coordination relative to STS and Tanpura listening.

AUM listening generated a well-organized network involving frontal, central, parietal, and occipital regions. Although the connectivity pattern was distributed and structured, the overall network density appeared lower than that observed during STS listening. The presence of several long-range fronto-occipital interactions suggests enhanced neural integration associated with the auditory processing of the chant.

Tanpura listening exhibited strong and highly coordinated connectivity characterized by multiple long-range interactions and elevated wPLI values. The resulting network was more extensive than the resting and MM conditions and approached the connectivity strength observed during STS listening, suggesting auditory entrainment and coordinated large-scale neural activity.

Collectively, these findings indicate that different auditory stimuli induce distinct patterns of functional brain organization. STS listening produced the most prominent large-scale synchronization, whereas MM elicited more localized connectivity. AUM and Tanpura listening were associated with organized distributed networks, reflecting enhanced neural coordination beyond the resting condition.

\section{Discussion}

The present pilot investigation explored the influence of different chant-listening conditions on neural oscillatory activity and functional brain organization using EEG. Although conducted on a single participant, the study provides preliminary insights into how rhythmic auditory stimulation may differentially modulate cortical activation and large-scale neural synchronization during early childhood.

The spectral analysis revealed that STS listening produced the highest power across multiple frequency bands, particularly within the beta range. Beta oscillations have been associated with attentional engagement, cognitive processing, and active information integration \cite{Klimesch1999,Engel2010}. The elevated spectral power observed during STS listening may therefore reflect enhanced attentional engagement and rhythmic entrainment induced by the dynamic acoustic structure of the chant. Previous studies have similarly reported increased cortical activation during structured chanting and rhythm-based auditory practices \cite{Lomas2015,Tang2015}.

Functional connectivity analysis further demonstrated marked differences across listening conditions. Among all conditions, STS listening exhibited the strongest and most widespread connectivity network, characterized by numerous high-strength long-range interactions linking frontal, temporal, parietal, and occipital regions. The predominance of strong wPLI connections suggests enhanced large-scale neural communication and synchronization, potentially reflecting increased attentional allocation and integrative processing elicited by the rhythmic and energetic nature of the stimulus.Similar increases in long-range functional connectivity have previously been reported during meditation and focused-attention practices, suggesting enhanced integration among distributed neural systems \cite{jang2011increased}.

Tanpura listening also produced a relatively dense connectivity pattern with several high-strength long-range interactions. Unlike STS, however, the connectivity appeared more balanced and less globally distributed. Continuous harmonic stimulation from the Tanpura may facilitate neural entrainment mechanisms that promote coordinated oscillatory activity while maintaining a relaxed cognitive state. Auditory entrainment studies have shown that rhythmic and harmonic acoustic stimulation can synchronize neural oscillations and strengthen large-scale cortical communication \cite{thaut2015neural}. Similar effects of rhythmic auditory stimulation on cortical synchronization have been reported in auditory entrainment studies \cite{Nozaradan2012}.

AUM listening generated a moderately organized network characterized by interactions involving frontal, central, and posterior regions. While the overall connectivity density was lower than that observed during STS listening, the presence of distributed inter-regional connections suggests coordinated neural processing associated with sustained auditory attention and contemplative engagement. Previous studies have reported alterations in neural synchronization during exposure to mantra-based and meditative auditory stimuli \cite{Kalyani2011,Lomas2015}.

In contrast, MM listening demonstrated comparatively weaker and more localized connectivity patterns. The predominance of lower-strength connections suggests reduced large-scale synchronization relative to STS and Tanpura conditions. Nevertheless, selective interactions involving frontal and posterior regions indicate that the stimulus still engaged distributed neural networks to a certain extent.

The broader significance of this work extends beyond the immediate investigation of chant listening. One of the primary motivations underlying this study is the development of objective neurophysiological markers capable of characterizing cognitive and developmental trajectories in children. India currently lacks large-scale longitudinal neurodevelopmental datasets integrating neural, cognitive, behavioral, and educational measures. Consequently, there remains limited understanding of how neural development interacts with educational environments, cultural practices, language diversity, and socioeconomic factors across the country.

Emerging evidence suggests that EEG-derived biomarkers can provide valuable insights into developmental processes and neurodevelopmental disorders such as Autism Spectrum Disorder (ASD), Attention Deficit Hyperactivity Disorder (ADHD), dyslexia, and specific learning disabilities \cite{Jeste2015,Uddin2017}. Alterations in neural oscillations, connectivity patterns, and network organization have been consistently reported across these conditions. Establishing normative developmental EEG profiles may therefore facilitate early identification of atypical developmental trajectories and support timely intervention.

Chant listening, rhythmic auditory stimulation, and contemplative practices may be particularly relevant in this context. Neuroplasticity research indicates that repeated attentional and sensory training can modify functional brain networks and improve cognitive regulation \cite{Davidson2003,Lutz2004}. Such culturally relevant and low-cost interventions may offer promising complementary approaches for supporting cognitive development and educational outcomes in children.

Several limitations should be acknowledged. First, the investigation involved a single participant and therefore does not permit statistical inference. Second, only sensor-level spectral and connectivity analyses were performed. Third, behavioral and cognitive measures were not incorporated. Future studies should include larger cohorts, longitudinal assessments, source-space analyses, graph-theoretical metrics, and neurocognitive evaluations to better understand the relationship between auditory interventions and developmental brain function.

Despite these limitations, the present work establishes a methodological framework for investigating chant-induced neural dynamics and demonstrates that different auditory conditions engage distinct patterns of cortical activation and functional connectivity. These findings support the feasibility of EEG-based approaches for studying neuroplasticity, cognitive development, and potential intervention strategies in child-centered neuroeducation research.

\section{Conclusion}

This pilot EEG investigation examined neural activity across five auditory listening conditions using spectral and functional connectivity analyses. Distinct neural signatures were observed for each condition, highlighting the sensitivity of EEG measures to variations in rhythmic and contemplative auditory stimulation.

STS listening produced both the highest spectral power and the strongest functional connectivity network, suggesting enhanced cortical activation and large-scale neural synchronization. Tanpura listening generated a dense yet balanced connectivity pattern consistent with auditory entrainment and coordinated neural processing. AUM listening exhibited moderate distributed connectivity, whereas MM listening demonstrated comparatively weaker and more localized network organization. Resting-state recordings showed lower overall synchronization relative to the active auditory conditions.

These findings suggest that different forms of chant listening engage distinct neural mechanisms involving both oscillatory activity and inter-regional communication. In particular, rhythmic and structured auditory stimuli such as STS appear capable of enhancing large-scale functional integration, while sustained harmonic stimulation such as Tanpura listening may support coordinated yet relaxed neural states.

Beyond the immediate findings, this study represents an initial step toward developing EEG-based developmental biomarkers for educational neuroscience and child development research in India. Understanding how culturally relevant auditory practices influence neural activity may contribute to future efforts aimed at characterizing neurotypical developmental trajectories, identifying children who may benefit from early intervention, and designing neuroscience-informed educational and therapeutic strategies.

Future work will involve larger longitudinal cohorts, multimodal cognitive assessments, and advanced neurophysiological analyses to evaluate whether chant-listening interventions can facilitate cognitive development, attention regulation, and neurorehabilitation in both neurotypical and neurodivergent populations.
\section*{Acknowledgment}

The authors would like to acknowledge the support and guidance provided by the National Institute of Advanced Studies (NIAS), Bengaluru, and the Centre for Brain Research (CBR), Indian Institute of Science (IISc), Bengaluru.

\bibliographystyle{IEEEtran}
\bibliography{refs}

\end{document}